\documentclass[preprint2]{aastex}

\usepackage{rotating}

\newcommand{\scs}{\scriptsize}
\newcommand{\targ}{4U\thinspace1626-67}
\newcommand{\hst}{{\em HST}}
\newcommand{\cxo}{{\em Chandra}}

\newcommand{\til}{$\sim$}
\newcommand{\lam}{$\lambda$}
\newcommand{\msun}{\thinspace\hbox{$M_{\odot}$}}
\newcommand{\ergsqcmsec}{\thinspace\hbox{$\hbox{erg}\thinspace\hbox{cm}^{-2}
                \thinspace\hbox{s}^{-1}$}}
\newcommand{\ergsec}{\thinspace\hbox{$\hbox{erg}\thinspace\hbox{s}^{-1}$}}

\newcommand{\simgt}{\mbox{$^{>}\mskip-10.5mu_\sim$}}

\newcommand{\kms}{\thinspace\hbox{$\hbox{km}\thinspace\hbox{s}^{-1}$}}

\def\today{\ifcase\month\or
January\or February\or March\or April\or May\or June\or
July\or August\or September\or October\or November\or December\fi
\space\number\day, \number\year}

\slugcomment{Accepted for publication by The Astronomical Journal}

\shorttitle{UV Spectrum of the X-ray binary-- \targ}
\shortauthors{Homer et al.}

\begin{document}


\title{The UV Spectrum of the Ultra-compact X-ray Binary-- \targ\footnote{\ Based on observations with
the NASA/ESA
Hubble Space Telescope, obtained at the Space Telescope Science Institute,
which is operated by the Association of Universities for Research in
Astronomy, Inc., under NASA contract NAS5-26555.}
}

\author{L. Homer, Scott F. Anderson and Stefanie Wachter}
\affil{Astronomy Department, Box 351580, University of Washington, Seattle, WA 98195-1580}
\email{homer,anderson,wachter@astro.washington.edu}
\and
\author{Bruce Margon}
\affil{Space Telescope Science Institute, 3700 San Martin Drive, Baltimore, MD 21218}
\email{margon@stsci.edu}
\author{\sl \small Accepted for publication by The Astronomical Journal}




\begin{abstract}
We have obtained {\em Hubble Space Telescope}/STIS low-resolution ultraviolet
spectra of the X-ray pulsar \targ\ (=KZ TrA); \targ\ is unusual
even among X-ray pulsars due to its ultra-short binary period (P=41.4 min)
and remarkably low mass-function ($\leq1.3\times10^{-6}$\msun).
The far-UV spectrum was exposed for a total of 32ks and has
sufficient signal-to-noise to reveal numerous broad emission and prominent
narrower absorption lines.  Most of the
absorption lines are consistent in strength with a purely interstellar
origin.  However, there is evidence that both \ion{C}{1} and \ion{C}{4}
require additional absorbing gas local to the system. In emission,
the usual prominent lines of \ion{N}{5} and \ion{He}{2} are absent, whilst both
\ion{O}{4} and \ion{O}{5} are relatively strong.   We further identify a 
rarely seen feature at \til1660\AA\ as the \ion{O}{3}] multiplet. Our
ultraviolet spectra therefore provide independent support for the recent suggestion that the mass 
donor is the chemically fractionated core of either a C-O-Ne or O-Ne-Mg white 
dwarf; this was put forward to explain the results of \cxo\  high-resolution X-ray spectroscopy.  The velocity profiles of the ultraviolet lines are in all
cases broad and/or flat-topped, or perhaps even double-peaked 
for the highest ionization cases of O; in either case the ultraviolet 
line profiles are in broad 
agreement with the Doppler pairs found in the X-ray spectra.  Both the X-ray 
and far-UV lines are plausibly formed in (or in an corona just above) a 
Keplerian accretion disc; the combination of ultraviolet and X-ray spectral 
data may provide a rich data set for follow-on detailed models of the disk 
dynamics and ionization structure in this highly unusual low-mass X-ray pulsar 
system.

\end{abstract}


\section{INTRODUCTION}
The X-ray source \targ\ is one of the rare cases in which an 
X-ray pulsar is a member of a low mass X-ray binary (LMXB) system. Its 7.7s pulsations were first detected in 1977 by {\it SAS-3} \citep{rapp77}, but even with
further extensive observations the pulse timing has yet to reveal any indication of a binary orbit, placing very tight constraints on the mass function
\citep[$\leq1.3\times10^{-6}$\msun,][]{levi88}.  However, following the identification of a
faint, blue optical counterpart (KZ TrA) by \citet{Mcc77}, extensive fast optical photoelectric photometry provided convincing evidence for binarity.  In addition to
the direct optical pulses due to the reprocessing by the disc of the pulsed X-ray flux, \citet{midd81} uncovered the Doppler-shifted signal from
optical pulses due to reprocessing on the donor star's heated face.  This detection has now been confirmed (with the same instrumentation) by
\citet{chak98b}, and recently with a fast frame-transfer CCD camera \citep{chak01}. 

The derived orbital period of 41.4 min places \targ\ as one of the
ultra-short period LMXBs, i.e. those with  $P_{orb}<80$ min, below the limit for a low-mass main sequence hydrogen-burning donor.  In
consequence there has been much controversy regarding the nature of the secondary and the evolutionary history of
the system. The earliest proposed
donor types were a low-mass (0.08\msun) severely hydrogen-depleted  and partially degenerate star \citep{nels86}, or an even lower mass
(\til0.02\msun) helium or C-O white dwarf \citep{verb90}.  Later moderate resolution X-ray spectra from {\it ASCA} \citep{ange95} and {\it
BeppoSAX} \citep{owen97} revealed emission
line structures identified as due to Ne/O, indicating an overabundance of these elements in the system. To explain the excess Ne (a
by-product of He burning) \citet{ange95} proposed that the donor could be a 
a low-mass helium burning star, under-filling its Roche lobe and transferring material via a powerful stellar wind. The most recent progress has come from high resolution X-ray grating spectroscopy from
\cxo.  \citet{schu01} (hereafter SCH01) report on further abundance anomalies required to explain the strength of various absorption edges. From the inferred
local abundance ratios they now argue that the mass donor is most
likely the 0.02\msun\ chemically fractionated core of a {C-O-Ne} or O-Ne-Mg white dwarf. An evolutionary scenario has also been developed by
\citet{yung02}, indicating a probable 0.3--0.4\msun\ C-O white dwarf as progenitor.

The \cxo\ spectra not only resolve the individual X-ray emission lines, but they also indicate double-peaked emission, especially in the case of \ion{Ne}{10} \lam12. SCH01 interpret these as Doppler pairs arising most probably in or near the Keplerian disc
flow. Indeed, this would be inconsistent with a more massive 0.08\msun\ donor, for which $i\leq  8^{\circ}$;  whereas for a 0.02\msun\
white dwarf the maximum allowed $i=30^{\circ}$ makes it much more plausible.

\targ\ has now been studied extensively in the X-ray regime and also
photometrically in the optical.  However, little useful spectral information
has been extracted in the optical,
as LMXBs generally show only a
few weak lines there. \citet{cowl88} obtained two spectra of \targ\ which 
show only weak \ion{C}{3}/\ion{N}{3} 4650/60 emission. In the far-ultraviolet (FUV), however, the spectra
of LMXBs are normally characterized by numerous strong emission lines 
due to the ionizing flux of the incident X-rays.  Hence, we have utilized the excellent UV capability of the Space Telescope Imaging
Spectrograph \citep[STIS,][]{wood98} on-board the {\it Hubble Space Telescope (HST)} to obtain time-resolved spectra of \targ\ in both this 
line-rich FUV region and also in the near-UV (NUV).  In this paper, we report our results on the
time-averaged spectra, concentrating on the FUV lines, and discuss our findings in the light of the recent X-ray spectral results in particular.

\begin{figure*}[!htb]
\resizebox{1.05\textwidth}{!}{\rotatebox{-90}{\plotone{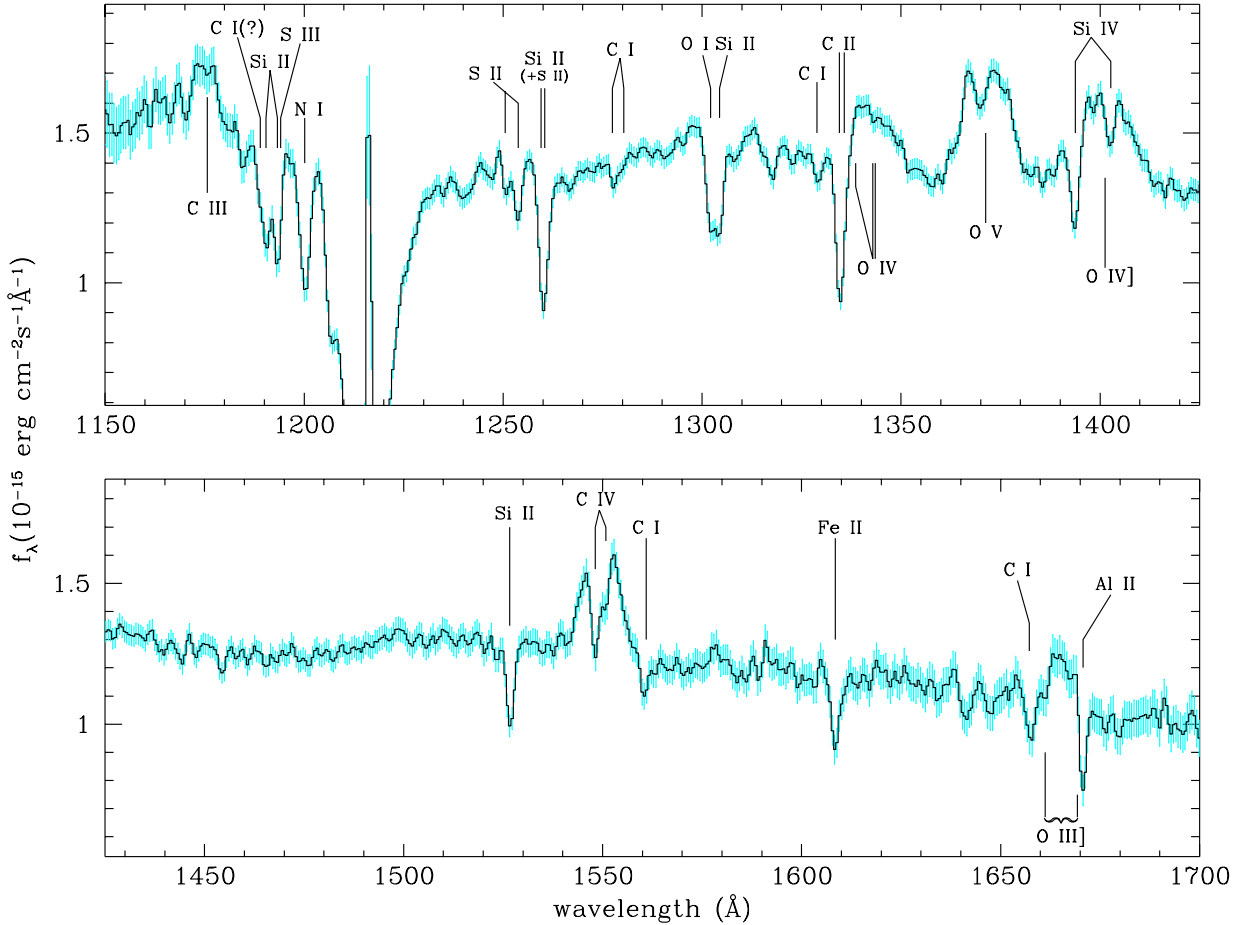}}}
\caption{Far-UV spectrum of \targ\ (=KZ TrA) from a total exposure of 31.2 ks with \hst/STIS. No correction for the small reddening of this object has
been made, but to aid clarity 3 pixel boxcar smoothing has been
applied. Numerous emission and absorption lines are present, the most significant of which have been labelled. Emission lines are labelled from
beneath, whilst absorption or coinciding emission/absorption lines are labelled from above. Note in particular the broad and
complicated emission profiles; these are discussed further in the text.\label{fig:fspec}}
\end{figure*}

\section{OBSERVATIONS AND DATA REDUCTION}
We obtained {\it HST}/STIS UV observations of \targ\ on 1999 April 2, May 31 and June 1 totaling 12 \hst\ orbits; we also extracted a further 4 orbits of archival data
taken a year earlier on 1998 April 23 as part of another program. All data were taken with either
the low resolution G140L (FUV) or G230L (NUV) gratings, which give useful wavelength ranges of \til1160--1700\AA\ and \til1700--3150\AA, at
1.2\AA\ and 3.2\AA\ 
resolutions respectively. The FUV and NUV MAMA
detectors were operated in Time-Tag mode, which is then effectively photon-counting\footnote{Time-resolved broad-band UV lightcurves are
presented in \citet{chak01}}. All these data were reduced
by the On-The-Fly Reprocessing system at STScI, which utilizes the best currently available
calibration files, software and data parameters.
The products provided include 2-D spectral images created by accumulating the photon event list over the exposure during a given \hst\ orbit, as well as the wavelength
and flux-calibrated 1-D spectra extracted from these. We
then also applied additional wavelength corrections using the STSDAS WAVECAL routine (if not previously completed), which makes use of the wavelength calibration (arc)
spectra taken with each science
exposure. Lastly, we derived a fully time-averaged spectra in each band from all the data (weighting each \hst\ orbit's data according to exposure time),
which amounts to total exposure times of 31.2 ks (8.7 hrs) in the FUV (and 10.8 ks in the NUV).

\section{DATA ANALYSIS AND RESULTS}
In figure~\ref{fig:fspec} we present the complete FUV spectrum (not corrected for the small reddening towards this object), with 3 pixel boxcar smoothing applied to aid clarity (and with 1$\sigma$ error-bars
combined accordingly).  The long exposure provides
good signal-to-noise, and numerous prominent broad emission lines and narrower absorption features stand out. By contrast, the NUV spectrum
shows no very
strong emission features and only absorption lines (from \ion{Mg}{2}, \ion{Mn}{2}, \ion{Fe}{2} and \ion{Zn}{2}) of strength consistent with
interstellar origin; hence we will only make occasional further reference to results from this NUV spectrum in the remainder of this paper.

Apart from the very strong absorption feature at \til1215\AA\ due to Ly$\alpha$, which has prominent damping wings, the results of Gaussian fits (using the full-resolution
data) to all the prominent FUV lines (measured line centers, FWHM and
equivalent widths) are presented in table~\ref{tab:lines_det}. Almost all of the emission lines exhibit complex structures and in most cases
emission and absorption lines coincide.  We will discuss the
fits to these more complex profiles in turn, in the following sections.  For the remaining absorption lines we fit each line with a single Gaussian, with
center, width and normalization as free parameters. In the case of blends, if the separation of the two (or more) lines was greater than the
spectral resolution, we included a Gaussian profile for each.

Similarly for each line contributing to the emission/absorption line complexes we used unconstrained Gaussians.  However, for the components of multiplets additional constraints were applied. We required the ratio of the components' centers (effectively
the spacing) to agree with that of the vacuum wavelengths, and the FWHM of each component to be the same. The absorption components are likely
dominated by interstellar absorption.  Hence we left the ratio of the
fluxes (e.g. the doublet ratio) free to allow for the probable range of optical depth for the absorbing clouds.  However, attempts at leaving the
ratios free for the emission too, indicated that the data are insufficient to meaningfully constrain these values, and often actually
produced unphysical results.  Hence, in most cases we simply assumed the optically thick limit for the emission lines and used the
appropriate relative intensities as tabulated in the National Institute of Standards and Technology (NIST) listings\footnote{available at
http://physics.nist.gov/cgi-bin/AtData/lines\_form}. Most observational
determinations, as well as
the results of modeling, \citep[see e.g.][]{ko94}, indicate that these resonant emission lines are typically optically
thick. Nevertheless, for the \ion{Si}{4} and \ion{C}{4} doublets we also examined the optically thin limit
(where the doublet ratio is 2:1 for the lower to higher wavelength components), but found no substantive differences (see the relevant section for
details). We therefore tabulate (and show figures) based on the optically thick assumption. Lastly, we also tried
twin-Gaussians to represent two separate blue- and red-shifted emission regions, i.e. as was done for the double-peaked X-ray lines by SCH01.

\renewcommand{\arraystretch}{.75}
\begin{deluxetable}{l l l l c c l l }
\tablewidth{0pt}
\tablecaption{Parameters of Gaussian fits to FUV absorption and emission lines in \targ \label{tab:lines_det}}
\tablehead{
\colhead{Compo}& 
\colhead{ID     } &
\colhead{$\lambda_{vac}$}&
\colhead{$\Delta\lambda$}&
\colhead{$V$  } & 
\colhead{ FWHM  } & 
\colhead{ Flux ($10^{-16}$ } & 
\colhead{EW  }\\

\colhead{-nent\tablenotemark{a}} & 
\colhead{	} &
\colhead{	} & 
\colhead{ 		(\AA)		} & 
\colhead{	(\kms) } & 
\colhead{(\kms) } & 
\colhead{ \ergsqcmsec) } & 
\colhead{(\AA)			 }\\
}
\startdata
+&\ion{C}{3}   &1175.64\tablenotemark{b} &$-0.33\pm0.52 $&$ -85\pm 135$&$ 1440\pm  330$&\phs$ 10.6 \pm 3.4  $&$0.68 \pm 0.22 $\\[1.9mm]

--&\ion{C}{1}(?)  &1188.99 &$-0.11\pm0.25 $&$ -30\pm  65$&$  380\pm  150$&\phn$ -3.9 \pm 1.6  $&$ 0.26 \pm 0.11 $\\
--&\ion{Si}{2}  &1190.42 &\phs$0.25\pm0.11 $&\phs$65\pm  30$&$  360\pm   70$&\phn$ -6.5 \pm 1.4  $&$ 0.45 \pm 0.10 $\\
--&\ion{Si}{2}\tablenotemark{c}   &1193.29 &$-0.01\pm0.06 $&\phn$  -5\pm  15$&$  470\pm   40$&\phn$ -9.2 \pm 1.0  $&$ 0.63 \pm 0.07 $\\
--&+\ion{S}{3}\tablenotemark{c}  &1194.06 &$-0.78\pm0.06 $&$-195\pm  15$&	&		&		\\[1.9mm]

--&\ion{N}{1}\tablenotemark{b}     &1200.13 &\phs$ 0.16\pm0.06 $&\phs$  40\pm  15$&$  610\pm   40$&$-12.3 \pm 1.0  $&$ 0.88 \pm 0.07 $\\[1.9mm]
--&\ion{S}{2}    &1250.58 &$-0.98\pm0.40 $&$-235\pm  95$&$  850\pm  230$&$-10.5 \pm 2.8  $&$ 0.76 \pm 0.20 $\\

--&\ion{S}{2}    &1253.81 &$-0.02\pm0.11 $&\phs$  -5\pm  25$&$  460\pm   90$&\phn$ -5.9 \pm 1.3  $&$ 0.42 \pm 0.09 $\\[1.9mm]
--&\ion{Si}{2}\tablenotemark{c}   &1260.42 &$-0.28\pm0.05 $&$ -65\pm  10$&$  560\pm   30$&$-12.6 \pm 0.9  $&$ 0.93 \pm 0.07 $\\

--&(+\ion{S}{2})\tablenotemark{c} &1259.52 &\phs$ 0.62\pm0.05 $&\phs$ 150\pm  10$&	&		&		\\[1.9mm]
--&\ion{C}{1}     &1277.41 &\phs$ 0.33\pm0.21 $&\phs$  75\pm  50$&$  320\pm  110$&\phn$ -1.9 \pm 0.8  $&$ 0.13 \pm 0.06 $\\

--&\ion{C}{1}     &1280.33 &$-0.54\pm0.80 $&$-125\pm 190$&$  380\pm  280$&\phn$ -1.2 \pm 1.0  $&$ 0.09 \pm 0.07 $\\[1.9mm]
--&\ion{O}{1}     &1302.17 &\phs$ 0.04\pm0.14 $&$  10\pm  30$&$  600\pm  110$&$-12.0 \pm 2.7  $&$ 0.81 \pm 0.19 $\\

--&\ion{Si}{2}   &1304.37 &\phs$ 0.14\pm0.11 $&$  30\pm  25$&$  390\pm   70$&\phn$ -6.8 \pm 1.9  $&$ 0.46 \pm 0.13 $\\[1.9mm]

--&\ion{C}{1}     &1328.83 &\phs$ 0.44\pm0.23 $&$ 100\pm  50$&$  530\pm  170$&\phn$ -2.9 \pm 1.0  $&$ 0.20 \pm 0.07 $\\

--&\ion{C}{2}    &1334.53 &$-0.01\pm0.16 $&\phs$   0\pm  35$&$  500\pm   50$&$-12.0 \pm 1.9  $&$ 0.85 \pm 0.14 $\\

--&\ion{C}{2}   &1335.71 &$-0.01\pm0.16 $&\phs$   0\pm  35$&$  500\pm   50$&\phn$ -4.6 \pm 2.4  $&$ 0.32 \pm 0.17 $\\

+&\ion{O}{4}   &1338.61 &\phs$ 0.37\pm0.47 $&\phn$  85\pm 105$&$ 2780\pm  380$&\phn\phs$  9.7 \pm 1.4  $&$0.69 \pm 0.10 $\\

+&\ion{O}{4}   &1342.99 &\phs$ 0.37\pm0.47 $&\phn$  85\pm 105$&$ 2780\pm  380$&\phn\phs$  6.5 \pm 1.0  $&$0.43 \pm 0.07 $\\

+&\ion{O}{4}   &1343.51 &\phs$ 0.37\pm0.47 $&\phn$  85\pm 105$&$ 2780\pm  380$&\phn\phs$  9.1 \pm 1.4  $&$0.66 \pm 0.10 $\\[1.9mm]

--&\ion{C}{1}     &1328.83 &\phs$ 0.30\pm0.18 $&\phs$  70\pm  40$&$  440\pm  100$&\phn$  -2.2 \pm 0.7  $&$  0.15 \pm 0.05 $\\

--&\ion{C}{2}    &1334.53 &$-0.21\pm0.20 $&$ -45\pm  45$&$  400\pm   70$&\phn$  -9.1 \pm 1.9  $&$  0.64 \pm 0.13 $\\

--&\ion{C}{2}   &1335.71 &$-0.21\pm0.20 $&$ -45\pm  45$&$  400\pm   70$&\phn$  -6.6 \pm 3.0  $&$  0.47 \pm 0.21 $\\

+{\em b$^\ast$}&\ion{O}{4}   &1338.61 &$-3.61\pm0.38 $&$-805\pm  85$&$  770\pm  230$&\phn\phs$   2.5 \pm 0.9  $&$ 0.18 \pm 0.07 $ \\
																																	  
+{\em b}$^\ast$&\ion{O}{4}   &1342.99 &$-3.61\pm0.38 $&$-805\pm  85$&$  770\pm  230$&\phn\phs$   1.6 \pm 0.6  $&$ 0.11 \pm 0.04 $ \\
																																	  
+{\em b}$^\ast$&\ion{O}{4}   &1343.51 &$-3.61\pm0.38 $&$-805\pm  85$&$  770\pm  230$&\phn\phs$   2.9 \pm 1.1  $&$ 0.21 \pm 0.08 $ \\
																																	  
+{\em r}$^\ast$&\ion{O}{4}   &1338.61 &\phs$ 2.88\pm0.77 $&\phs$ 645\pm 170$&$ 1850\pm  430$&\phn\phs$   6.2 \pm 1.6  $&$ 0.44 \pm 0.11 $ \\
																																	  
+{\em r}$^\ast$&\ion{O}{4}   &1342.99 &\phs$ 2.88\pm0.77 $&\phs$ 645\pm 170$&$ 1850\pm  430$&\phn\phs$   4.0 \pm 1.0  $&$ 0.27 \pm 0.07 $ \\
																																	  
+{\em r}$^\ast$&\ion{O}{4}   &1343.51 &\phs$ 2.88\pm0.77 $&\phs$ 645\pm 170$&$ 1850\pm  430$&\phn\phs$   7.1 \pm 1.8  $&$ 0.52 \pm 0.13 $ \\[1.9mm]

+&\ion{O}{5}     &1371.29 &\phs$ 0.27\pm0.31 $&\phs$  60\pm  70$&$ 2820\pm  190$&\phs$  49.1 \pm 4.3  $&$ 3.66 \pm 0.32 $\\*[1.9mm]

+{\em b}$^\ast$&\ion{O}{5}     &1371.29 &$-4.49\pm0.22 $&$-980\pm  50$&$  950\pm  150$&\phs$ 14.2 \pm 2.6  $&$1.04 \pm 0.19 $\\*

+{\em r}$^\ast$&\ion{O}{5}     &1371.29 &\phs$ 2.75\pm0.30 $&\phs$ 600\pm  65$&$ 1680\pm  150$&\phs$ 28.9 \pm 3.0  $&$2.12 \pm 0.22 $\\*[1.9mm]

+&\ion{O}{5}     &1371.29 &\phs$ 0.20\pm0.18 $&\phs$  45\pm  40$&$ 1690\pm   70$&\phs\phd$ 491   \pm20    $&$36.1  \pm 1.5  $\\*

--&\ion{O}{5}     &1371.29 &$-0.21\pm0.19 $&$ -45\pm  40$&$ 1610\pm   80$&\phd$-447   \pm21    $&$ 32.9  \pm 1.6  $\\*[1.9mm]

+&\ion{Si}{4}   &1393.76 &\phs($\lambda$ fixed)&---&$  850\pm  460$&\phs$ 12.5 \pm12.2  $&$0.93 \pm 0.91 $\\*
--&\ion{Si}{4}  &1393.76 &\phs($\lambda$ fixed)&---&$  590\pm  130$&$-18.4 \pm8.0  $&$ 1.36 \pm 0.59 $\\*

+&\ion{O}{4}]  &1401.35 &\phs($\lambda$ fixed)&---&$ 3060\pm  180$&\phs$ 42.5 \pm4.5  $&$3.17 \pm 0.34 $\\*

+&\ion{Si}{4}  &1402.77 &\phs($\lambda$ fixed)&---&$  840\pm  450$&\phs$ 10.1 \pm9.9  $&$0.76 \pm 0.74 $\\*

--&\ion{Si}{4}  &1402.77 &\phs($\lambda$ fixed)&---&$  590\pm  130$&$-12.0 \pm5.2  $&$ 0.90 \pm 0.39 $\\[1.9mm]

+&\ion{Si}{4}   &1393.76 &\phs$ 2.79\pm0.37 $&$ 600\pm  80$&$ 2000\pm  160$&\phs$ 24.3 \pm 2.8  $&$1.81 \pm 0.21 $\\*

--&\ion{Si}{4}  &1393.76 &$-0.02\pm0.08 $&$  -5\pm  15$&$  520\pm   50$&$-10.3 \pm 1.2  $&$ 0.76 \pm 0.09 $\\*

+&\ion{O}{4}]  &1401.35 &\phs$ 1.52\pm0.27 $&$ 325\pm  60$&$  680\pm   10$&\phs$ 13.4 \pm 8.4  $&$1.00 \pm 0.63 $\\*

+&\ion{Si}{4}  &1402.77 &\phs$ 2.81\pm0.37 $&$ 600\pm  80$&$ 1990\pm  160$&\phs$ 19.4 \pm 3.7  $&$1.46 \pm 0.28 $\\*

--&\ion{Si}{4}  &1402.77 &$-0.02\pm0.08 $&$  -5\pm  15$&$  520\pm   50$&$-13.8 \pm 1.3  $&$ 1.03 \pm 0.10 $\\[1.9mm]
&& &&&&&\\*
		&& &&&&&\\*
		&& &&&&&\\*
		&& &&&&&\\*
		&& &&&&&\\*
		&& &&&&&\\*
		&& &&&&&\\*
		&& &&&&&\\*
		&& &&&&&\\*
		&& &&&&&\\*
		&& &&&&&\\*
		&& &&&&&\\*	
	&& &&&&&\\
+$^\ast$&\ion{Si}{4}   &1393.76 &$-0.55 $\tablenotemark{d}& $-120          $\tablenotemark{d}&$  870           $\tablenotemark{d}&\phs$  13.2  $\tablenotemark{d}&$ 0.97$\tablenotemark{d}\\*
																																	 
--$^\ast$&\ion{Si}{4}   &1393.76 &$-0.23 $&$ -50          $&$  540           $&$ -13.9  $&$  1.04$\\*
																																	 
+{\em b}$^\ast$&\ion{O}{4}]   &1401.35 &$-3.77 $&$-805 $&$  770           $&\phs\phn$   8.8  $&$ 0.65$\\*
																																	 
+{\em r}$^\ast$&\ion{O}{4}]   &1401.35 &\phs$ 3.00 $&\phs$ 645$&$ 1850           $&\phs$  21.6  $&$ 1.63$\\*
																																	 
+$^\ast$&\ion{Si}{4}   &1402.77 &$-0.55 $&$-120          $&$  870           $&\phs$  10.5  $&$ 0.79 $\\*
																																	 
--$^\ast$&\ion{Si}{4}   &1402.77 &$-0.23 $&$ -50          $&$  540           $&\phn$  -9.5  $&$  0.71$\\[1.9mm]

--&\ion{Si}{2}   &1526.71 &\phs$ 0.11\pm0.07 $&$  20\pm  14$&$  380\pm   30$&\phn$ -7.2 \pm 0.8  $&$ 0.56 \pm 0.06 $ \\[1.9mm]

+&\ion{C}{4}    &1548.20 &\phs$ 0.26\pm0.15 $&$  50\pm  30$&$ 1490\pm  120$&\phs$ 27.5 \pm 5.2  $&$2.20 \pm 0.42 $ \\*
																																	 
--&\ion{C}{4}    &1548.20 &$-0.02\pm0.07 $&$  -5\pm  15$&$  430\pm   40$&$-14.2 \pm 1.9  $&$ 1.14 \pm 0.15 $ \\*
																																	 
+&\ion{C}{4}   &1550.77 &\phs$ 0.26\pm0.15 $&$  50\pm  30$&$ 1490\pm  120$&\phs$ 24.8 \pm 4.7  $&$1.98 \pm 0.37 $  \\*
																																	 
--&\ion{C}{4}   &1550.77 &$-0.02\pm0.07 $&$  -5\pm  15$&$  430\pm   40$&\phn$ -9.9 \pm 1.8  $&$ 0.79 \pm 0.14 $  \\*
																																	 
--&\ion{C}{1}    &1560.91 &$-0.46\pm0.17 $&$ -90\pm  35$&$  320\pm   90$&\phn$ -3.1 \pm 1.0  $&$ 0.25 \pm 0.08 $  \\[1.9mm]
																																	 
--&\ion{Fe}{2}   &1608.45 &\phs$ 0.00\pm0.13 $&\phs$   0\pm  25$&$  330\pm   80$&\phn$ -5.9 \pm 1.6  $&$ 0.51 \pm 0.13 $ \\[1.9mm]

--&\ion{C}{1}\tablenotemark{b}     &1657.22 &\phs$ 0.40\pm0.20 $&$  70\pm  35$&$  260\pm   90$&\phn$ -2.5 \pm 1.1  $&$ 0.23 \pm 0.10 $ \\*
																																	 
+&\ion{O}{3}]\tablenotemark{e} &1661.20-- &\phs$ 0.60\pm0.30 $&$ 110\pm  55$&\phn$  580\pm  110$&\phs$ 17.1 \pm 4.6  $&$1.63 \pm 0.44 $ \\*
																																		 
&\scs{(7 comps)}&--1669.30&&& &&\\*																						 
																																	 
--&\ion{Al}{2}  &1670.79 &$-0.24\pm0.07 $&$ -45\pm  15$&$  280\pm   30$&\phn$ -9.5 \pm 1.5  $&$ 0.91 \pm 0.15 $  \\
\enddata
\tablecomments{The entries in the table are grouped (separated by horizontal spaces) according to which lines were blended together and
therefore required simultaneous fitting.}

\tablenotetext{a}{Coding to clarify the various components, especially in the case of the more complex absorption/emission structures: +
indicates emission component; -- indicates absorption component; when two emission components (double-peaked case) are present {\em r} and {\em
b} indicate
the red- and blue-shifted lines. Also, where more than one model has been fit to a profile, the $^\ast$ indicates the results of the preferred model.}
\tablenotetext{b}{Weighted average of vacuum wavelengths for closely spaced ($\Delta\lambda<1.2$\AA\ resolution) multiplet components quoted,
and single Gaussian used for fit}
\tablenotetext{c}{Single Gaussian used to fit blend of two lines with $\Delta\lambda<1.2$\AA\ spectral resolution.}
\tablenotetext{d}{No errors are quoted for the parameters for this fit, as they only represent a local minimum in $\chi^2$ space, that for
which the \ion{Si}{4} emission/absorption fluxes are smallest (and most plausible).}
\tablenotetext{e}{\ion{O}{3}] has 7 multiplet components in the range 1661.2\AA\ to 1669.3\AA\ as given by NIST, we used a Gaussian for each.  The
wavelength shift and FWHM apply to each component (constrained to be the same), whilst we quote the summed flux and EW only.}
\end{deluxetable}

\subsection{Fits to individual Emission/Absorption line complexes}
\subsubsection{\ion{C}{3} $\lambda1176$}
This profile appears to be flat topped, or even double-peaked.  However, given the lower signal-to-noise at this short wavelength end of the spectral range, such
details are well within the uncertainties and a simple single
Gaussian fit is as appropriate as any more complex model.  The resulting line center is consistent within the uncertainties with the weighted
average of the vacuum wavelengths
of the  closely spaced multiplet components.

\begin{center}
\leavevmode
\begin{figure*}[!htb]
\resizebox{.7\textwidth}{!}{\rotatebox{0}{\plotone{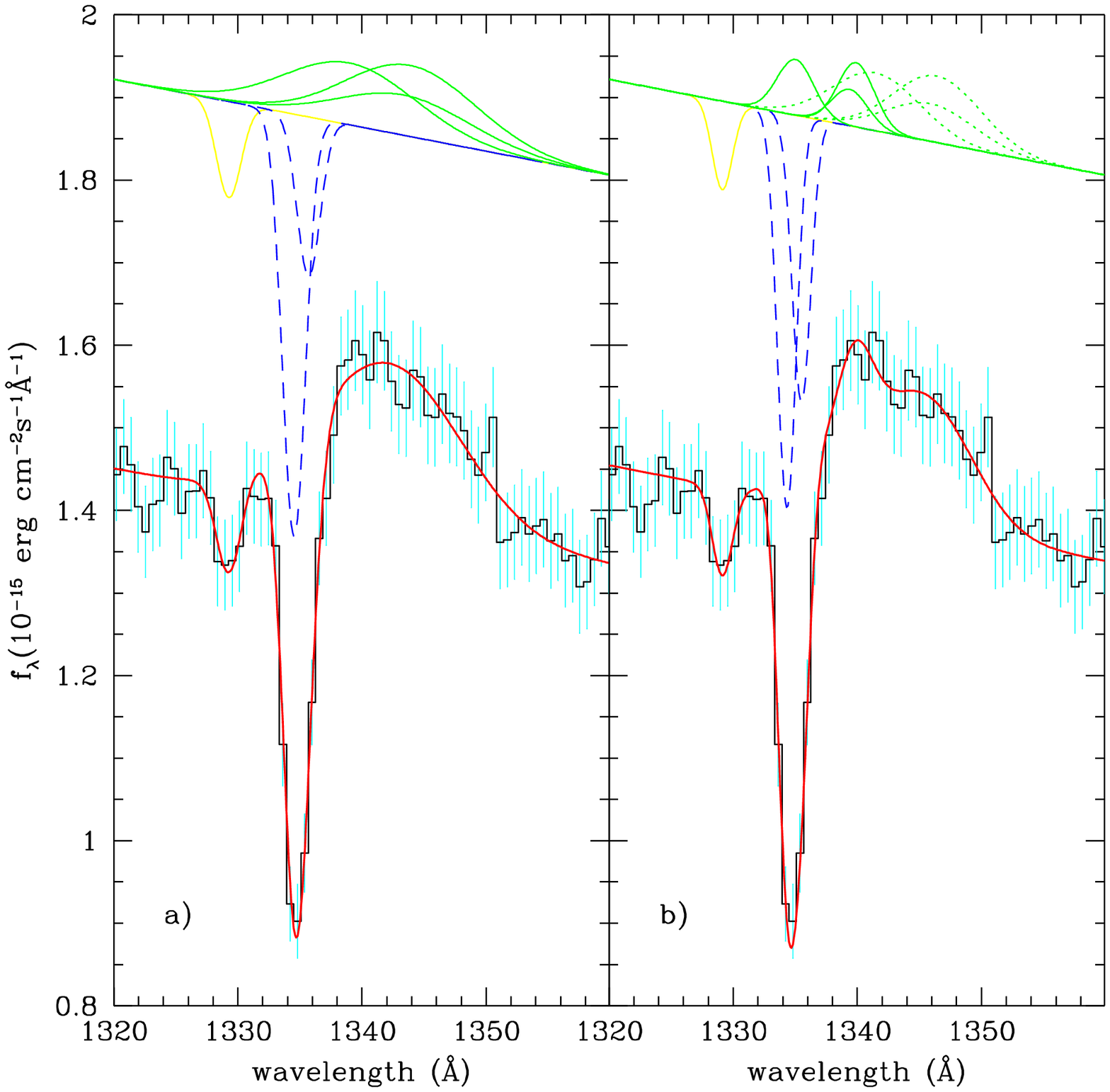}}}
\caption{Gaussian line profile fits to \ion{O}{4} triplet $\lambda1342$ emission, with absorption lines of \ion{C}{1} $\lambda1329$ (solid
line) and the \ion{C}{2} \lam1335,
\lam1336 doublet (dashed). a) Single emission component for each of the \ion{O}{4} triplet lines, note the very broad profile required (FWHM=2800\kms). b)
Assuming blue- (solid) and red-shifted (dotted) components for each triplet line provides a marginally better fit, but with significantly
broader red-shifted than blue-shifted lines.\label{fig:OIVfit}}
\end{figure*}
\end{center}

\subsubsection{\ion{O}{4} triplet $\lambda1342$}
In addition to the three emission components of the \ion{O}{4} triplet, one must also include absorption lines of \ion{C}{1} $\lambda1329$ and the \ion{C}{2}
\lam1335, \lam1336 doublet in order to fit the observed profile.  Even with the 1000\kms\ span of the three components of the multiplet, the breadth of this emission
feature requires a very large FWHM of 2800\kms\ for each separate component (see fig.~\ref{fig:OIVfit}a). A better representation may well be a two Gaussian double-peaked profile for each
component, as was used for the SCH01 X-ray line spectra. This more complex model does provide a better fit (fig.~\ref{fig:OIVfit}b), with a broad (FWHM=800\kms) blue-shifted component ($V=-800$\kms), but an even
broader (FWHM=1900\kms) red-shifted component ($V=600$\kms).

\subsubsection{\ion{O}{5} $\lambda1371$}
\begin{center}
\begin{figure*}[!htb]
\resizebox{.7\textwidth}{!}{\rotatebox{0}{\plotone{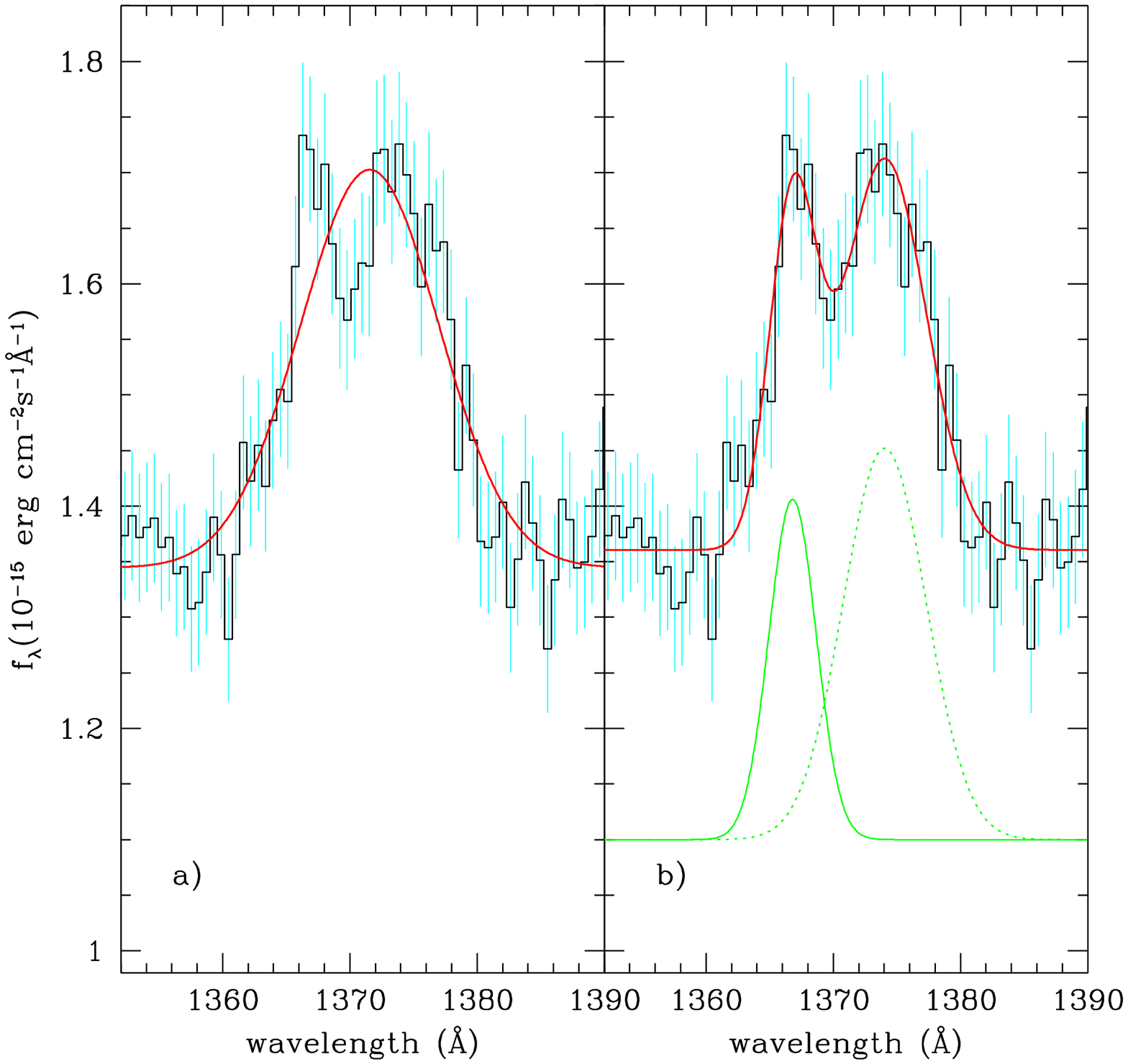}}}
\caption{Gaussian line profile fits to \ion{O}{5} $\lambda1371$. a) With a single emission
component a very broad line is required, but this is a noticeably poor fit due to the dip at \til1369\AA, b) double-peaked emission, with
red-shifted component (dotted line) much broader
than the blue (solid), providing a much better fit.\label{fig:OVfit}}
\end{figure*}
\end{center}

 This singlet line, unaffected by any expected strong interstellar absorption lines should be the most clear cut case.  However, the single Gaussian
 fit again yields a very large FWHM of $2800\kms$, as well as being clearly a very poor fit (see fig.~\ref{fig:OVfit}a). Once again motivated by the double-peaked X-ray lines, we have attempted such a fit, which works
 well (see fig.~\ref{fig:OVfit}b). This yields velocity shifts of --950\kms and 650\kms for the blue- and red-shifted components respectively, with reduced, but like \ion{O}{4}
 \lam1342, differing
 FWHM of 950 and
 1700 \kms.  For a disk representation the FWHM should be the same; hence this implies that there may be additional red-shifted emission
 contributions.  A fit using one blue-shifted and two red-shifted Gaussian
 components is also possible with a similar significance.

Our attempt to fit the profile with a single emission peak and absorption to account for the dip at \til1369\AA\ failed, although it is still possible that in
 addition to multiple emission peaks there is some absorption.  For instance, phase-resolved STIS FUV spectroscopy of
 Her~X-1 \citep{vrti01,boro00}  also revealed unusual features in the \ion{O}{5} line profile.  In addition to the broad and narrow
 emission components, absorption was required for the fits in Her X-1, which moved in phase with the binary and therefore was interpreted as a local
 effect.  We note that published STIS FUV spectra (phased-averaged) of the X-ray transient XTE J1859+226 \citep{hasw02} also shows a similar \ion{O}{5} structure
 to \targ, but
these authors  do not discuss the profile.

\subsubsection{\ion{Si}{4} \lam1394, \lam1403 doublet + \ion{O}{4}] $\lambda1401$ blend}
\begin{center}
\begin{figure*}[!htb]
\resizebox{1.0\textwidth}{!}{\rotatebox{-90}{\plotone{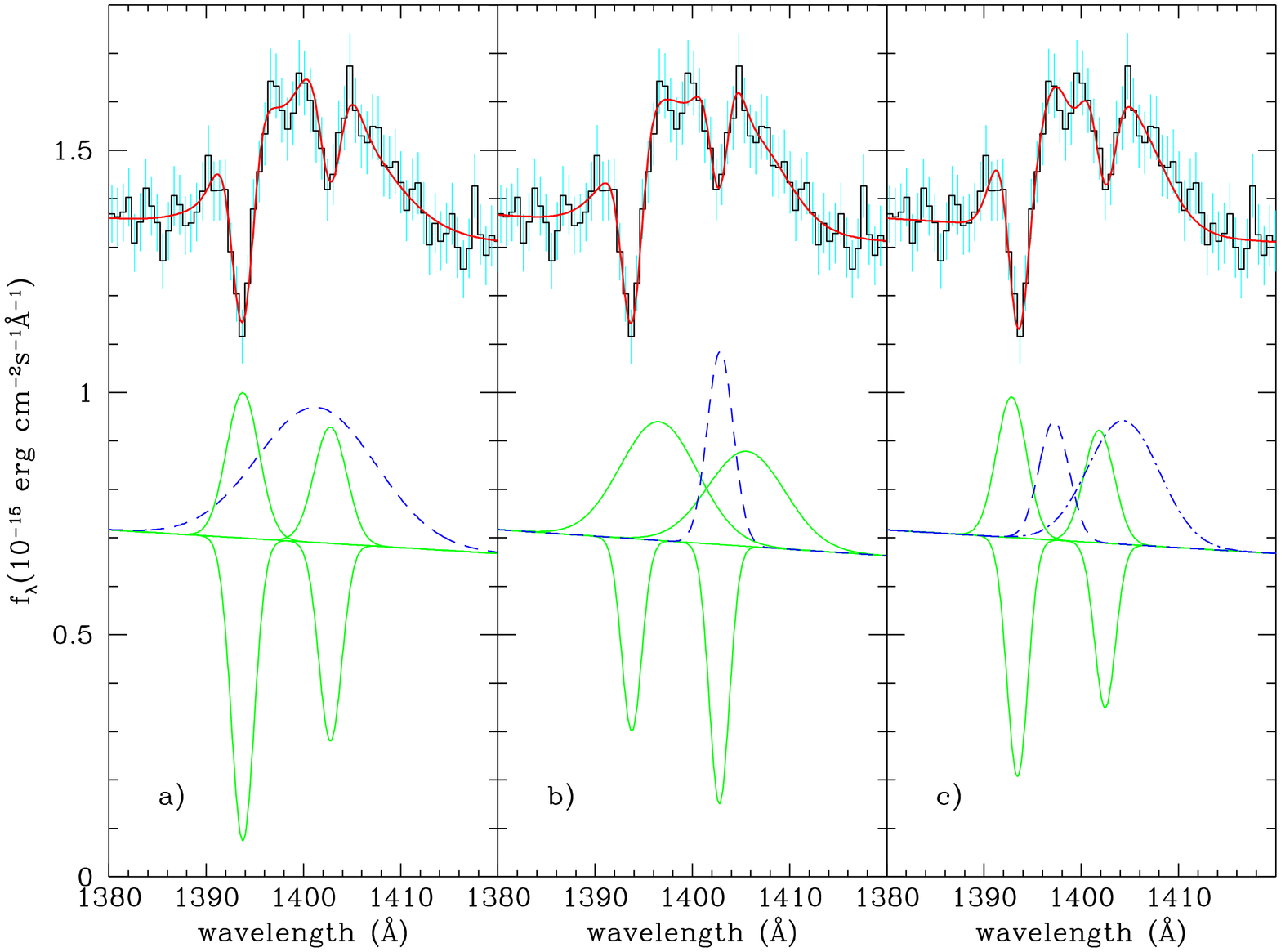}}}
\caption{Gaussian line profile fits to the complex Si~IV \lam1394, \lam1403 doublet (solid lines) + O~IV] $\lambda1401$ (dashed) blend. Both emission and absorption
components are included for the Si~IV doublet lines.  a) Constrained fit with line centers given by vacuum wavelengths.  This results in broad
and strong O~IV] emission, with narrow Si~IV emission, balancing only slightly narrower absorption lines.  b) With the line centers now
unconstrained (apart from the separation of the doublet components), the fit yields much broader Si~IV emission redshifted by as much as 600\kms, and narrow  O~IV] shifted to
balance the Si~IV \lam1403 absorption in an unlikely manner. c) Motivated by the fits to O~IV \lam1371 and O~V \lam1342, the single O~IV]
line has been replaced with a twin-Gaussian profile (dashed and dot-dashed), with separation, FWHM and relative normalizations constrained to that found for O~IV
\lam1342.  Essentially the results are similar to a), though the Si~IV emission is now slightly blue-shifted relative to its absorption.  None
of these fits is physically very satisfying, and in all likelihood the true underlying emission profiles are even more complex. \label{fig:SiIVfit}}
\end{figure*}
\end{center}

This is the most complex of the emission/absorption profiles in our spectrum.  Our most basic model is to assume broad emission components for
each of the \ion{Si}{4} doublet lines and the \ion{O}{4} line and narrower interstellar \ion{Si}{4} absorption, with the doublet ratio (DR) constrained
to the optically thick case (=1.25). We first
tried the model fixing all the central wavelengths to their vacuum values, and then relaxed this constraint.  In this instance, 
significant differences in the line parameters resulted. Although, in both cases the model can be made to fit the profile
quite well (apart from failing to delineate the drop in flux between the two \ion{Si}{4} lines at \til1398\AA), physical consideration of the
resulting parameters shows them to be questionable. The fits favor 
large line fluxes in both the emission and absorption components, which then balance out to provide the profile required.  In the fully
constrained fit almost all the emission flux comes from a very strong (EW=680\kms) and very broad (FWHM=3000\kms) \ion{O}{4}] line (see
  fig.~\ref{fig:SiIVfit}a), whilst in the less constrained fit a
large 600\kms\ velocity shift of the \ion{Si}{4} doublet is found, together with very high fluxes and 2000\kms\ line widths (see fig.~\ref{fig:SiIVfit}b).  Using a DR of 2 for the optically thin limit had little effect apart from the absorption intensities adjusting to a similar ratio to balance the
emission once again. Clearly much more complex emission components are needed. Since there is evidence that \ion{O}{5} and possibly \ion{O}{4} profiles are better
fit by an asymmetric twin-Gaussian model, we have tested whether such an \ion{O}{4} profile can fit the semi-forbidden line here.  We have
constrained the components to have the same velocity shifts, widths and relative normalizations as \ion{O}{4} \lam1342, and find that this does work
reasonably well (see fig.~\ref{fig:SiIVfit}c).  We note, however, that a similar double-peaked profile would be too broad to fit the
superimposed \ion{Si}{4} features as
they are relatively narrow.

\subsubsection{\ion{C}{4} doublet $\lambda1549$}
\begin{figure}[!htb]
\resizebox{.45\textwidth}{!}{\rotatebox{0}{\plotone{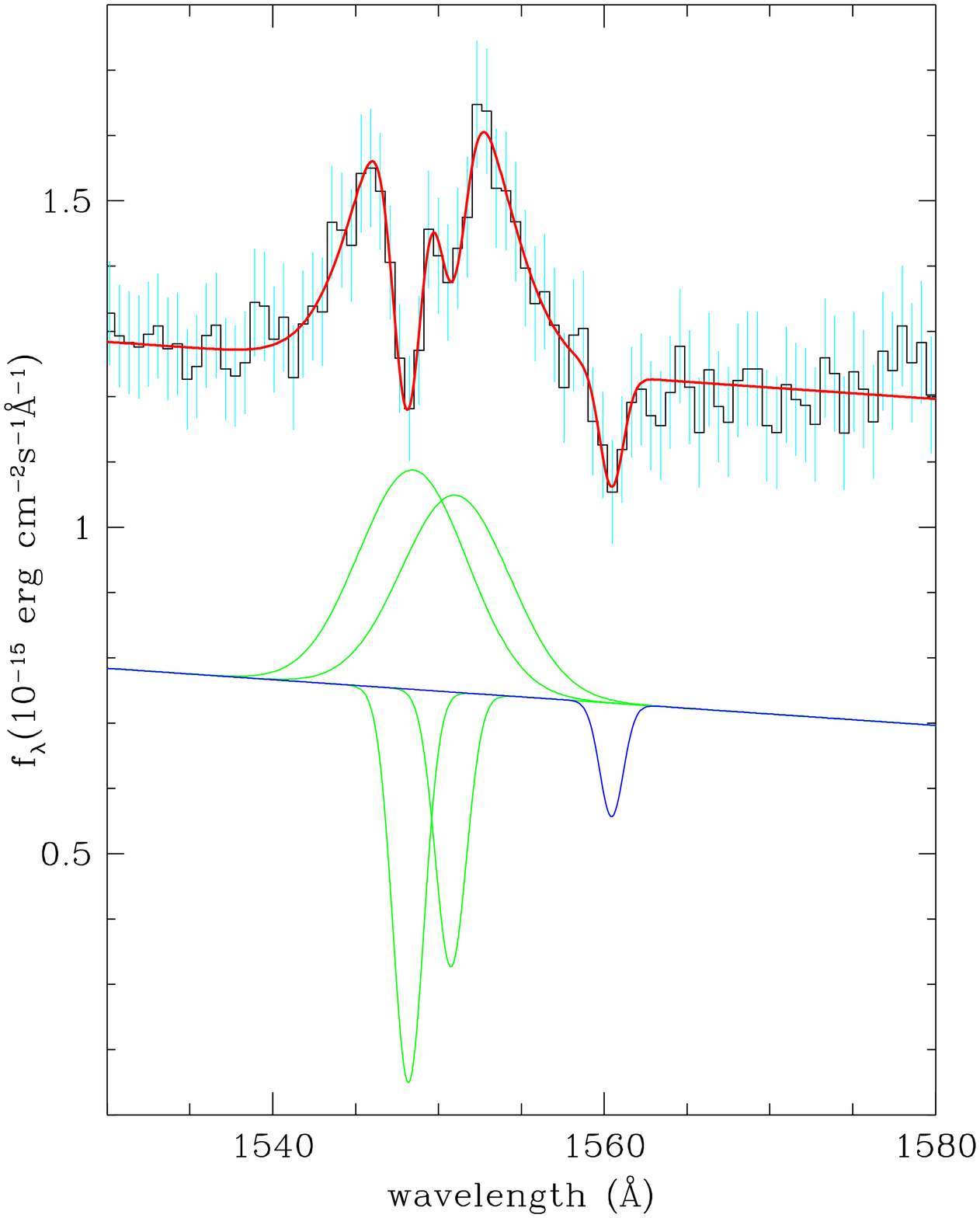}}}
\caption{Gaussian line profile fit to \ion{C}{4} doublet $\lambda1549$, including both emission and absorption, together with probable \ion{C}{1} \lam 1561
absorption. Fairly broad (FWHM=1500\kms) and strong \ion{C}{4} is required, with much narrower (FWHM=430\kms) absorption.  \label{fig:CIVfit}}
\end{figure}

As the baseline model for this doublet structure we assume the presence of the  two doublet emission lines (with DR=1.11 for optically thick
limit) and corresponding doublet
absorption lines, together with absorption at 1561\AA, which we attribute to \ion{C}{1}.  As can been seen in figure~\ref{fig:CIVfit}, this model does
provide a reasonable fit to the data, with only a small 50\kms\ shift (comparable to the uncertainty) in the emission doublet wavelength
required.  The strong \ion{C}{4} emission is a common feature of LMXB FUV spectra, but
the absorption is unusually pronounced.  Checking the optically thin limit for the emission (with DR=2), we found that the absorption strengths
are in fact insensitive to this ratio. Even trying double-peaked emission for each component of emission doublet cannot reduce the depth of the
absorption lines required, since the shape of the
line
profiles constrains each emission component to be almost as broad as the single Gaussians shown. We will discuss the strong \ion{C}{4} absorption and the probable \ion{C}{1} absorption further in \S\S~\ref{lab:ISA}.
Also, we note that the width/velocity spread of the emission lines is only about half that of the \ion{O}{5} and \ion{O}{4}.

\subsubsection{\ion{O}{3}] multiplet $\lambda1661-1669$}
This profile can be adequately fit with a single emission Gaussian for each of the multiplet components, together with  \ion{Al}{2} $\lambda1671$
and yet more \ion{C}{1} ($\lambda1657$) absorption. But, as for the \ion{O}{4} \lam1342 and \ion{O}{5} \lam1371 lines, a very large
FWHM\til2800\kms\ is indicated for each component by
the fit, and the feature is overall unusually strong for an LMXB, as compared to, for instance, the Si and C doublets.

\section{DISCUSSION}
\subsection{Absorption and emission strengths and the local elemental abundances} 
Henceforth, in the cases of the complex
emission and absorption profiles, we only consider the fluxes/EWs given by the physically reasonable fits.  These particular fits are indicated by asterisks in table~\ref{tab:lines_det}  to aid cross-reference.
 
\subsubsection{Interstellar versus local absorption}
\label{lab:ISA}
  At first glance, the equivalent widths (EWs) of the many absorption lines (if interstellar in origin) would appear rather large given the low
column towards this source, $E(B-V)=0.11\pm0.1$ (SCH01).  However, consideration of the results from the \hst\ Quasar Absorption Line Key
Project suggests otherwise for most of the lines. \citet{sava00} report on the Galactic interstellar features they identified in the spectra of this large
sample of 83 quasars.  Their sample includes a range of $E(B-V)$ from 0.014 to 0.155, as determined from radio \ion{H}{1} measurements;
therefore the range encompasses the reddening of \targ.  In table~\ref{tab:ISLcomp} we summarize our measured
equivalent widths for the numerous absorption lines (including those superimposed on top of emission), and present the comparative results from
\citet{sava00}, both their tabulated median and rms values for the EWs and an estimate of the maximum they measured (from their fig.~5). In almost all cases where a
comparison is possible, our measured values for \targ\ lie comfortably within the range for the quasars.  The only
exceptions are the \ion{Si}{4} \lam1393/1402, \ion{C}{2} \lam1335/1336 and  \ion{C}{4} \lam1548/1550. As noted above the \ion{Si}{4} \lam1394/1403 lines are
blended with \ion{O}{4}] making fitting to this profile rather uncertain.  Given the fact that the other Si lines (all \ion{Si}{2}) have EWs comparable
to typical interstellar absorption might suggest that the large \ion{Si}{4} EWs are an artefact of poor fitting and/or simply the use of too simplified a model. However, for C the situation is different, even ignoring
the complicated \ion{C}{4} doublet.  The \ion{C}{2} blend is too deep, though just consistent with the maximum interstellar EW value, and we also
identified a number of \ion{C}{1} absorption lines in our spectrum, of which only \ion{C}{1} \lam1329 is normally seen \citep[see e.g.][]{mort72}.  In the
case of \ion{C}{4} even trying double-peaked emission had very little effect on the amount of absorption required, and here the excess is
large.  The most likely explanation would seem to be excess C, at various ionization stages, local to the system.  Indeed, SCH01
found that fits to their X-ray spectra were improved by including the effects of a carbon $K$ edge with an overabundance of C relative to H of
$(5-18)\times$ solar.

\begin{deluxetable}{l l l l c c}[!htb]
\tablewidth{0pt}
\tablecolumns{6}
\tablecaption{Comparison of \targ\ Absorption line EWs to Interstellar lines from \hst\ Quasar Key Project \label{tab:ISLcomp}}
\tablehead{
\colhead{ ID }&
\colhead{ $\lambda_{vac}$ }&
\colhead{ Flux ($10^{-16}$ }&
\colhead{ EW }&
\multicolumn{2}{c}{EW from \hst\ Key Project\tablenotemark{a}}\\
\colhead{	}&
\colhead{	(\AA)			}&
\colhead{ \ergsqcmsec) }&
\colhead{ (\kms)}&
\colhead{Median (\kms) }&
\colhead{ Maximum (\kms)} \\
}

\startdata
\ion{C}{1}(?)  &1188.99 &\phn$ -3.88\pm 1.61 $&\phn$  67\pm  28$  & \phn-- & --\\
					  							
\ion{Si}{2}  &1190.42 &\phn$ -6.53\pm 1.39 $&$ 113\pm  24$  & \phn-- & --\\
					  							
\ion{Si}{2}   &1193.29 &\phn$ -9.20\pm 0.97 $&$ 159\pm  17$  & \phn-- & --\\
					  							
\ion{S}{3}  &1194.06 &\phn$ -9.20\pm 0.97 $&$ 159\pm  17$  & \phn-- & --\\
					  							
\ion{N}{1}     &1200.13 &$-12.28\pm 0.99 $&$ 219\pm  18$ & $240\pm65$ & 260\\
					  							
\ion{S}{2}    &1250.58 &$-10.49\pm 2.80 $&$ 182\pm  49$ & \phn-- & --\\
					  							
\ion{S}{2}    &1253.81 &\phn$ -5.91\pm 1.32 $&$ 101\pm  23$&\phn $70\pm90$ & 220\\
					  							
\ion{Si}{2}   &1260.42 &$-12.63\pm 0.94 $&$ 221\pm  18$& $190\pm65$ & 320\\
					  							
(+\ion{S}{2}) &1259.52 &&&&\\
					  							
\ion{C}{1}     &1277.41 &\phn$ -1.89\pm 0.79 $&\phn$  31\pm  13$ & \phn-- & --\\
					  							
\ion{C}{1}     &1280.33 &\phn$ -1.21\pm 0.99 $&\phn$  20\pm  16$ & \phn-- & --\\
					  							
\ion{O}{1}     &1302.17 &$-11.99\pm 2.73 $&$ 187\pm  43$& $105\pm55$& 220\\
					  							
\ion{Si}{2}   &1304.37 &\phn$ -6.84\pm 1.86 $&$ 106\pm  29$&\phn $95\pm35$& 190\\
					  							
\ion{C}{1}     &1328.83 &\phn$ -2.88\pm 1.03 $&\phn$  46\pm  16$ & \phn-- & --\\
					  							
\ion{C}{2}    &1334.53 &$-16.59\pm 3.10 $&$ 260 \pm  50 $& $160\pm45$ & 230\\
					  							
+\ion{C}{2}   &1335.71  &&&&\\
					  							
\ion{Si}{4}  &1393.76 &$-10.25\pm 1.17 $&$ 163\pm  19$&\phn$60\pm40$&160 \\
					  							
\ion{Si}{4}  &1402.77 &$-13.81\pm 1.32 $&$ 220\pm  21$&\phn$40\pm10$&60\\
					  							
\ion{Si}{2}   &1526.71 &\phn$ -7.17\pm 0.79 $&$ 109\pm  12$&\phn$95\pm25$&150\\
					  							
\ion{C}{4}    &1548.20 &$-14.21\pm 1.91 $&$ 220\pm  30$&\phn $80\pm25$&110\\
					  							
\ion{C}{4}   &1550.77 &\phn$ -9.92\pm 1.79 $&$ 153\pm  28$&\phn$45\pm20$&80\\
					  							
\ion{Fe}{2}   &1608.45 &\phn$ -5.87\pm 1.56 $&\phn$  94\pm  25$ & \phn-- & --\\
					  							
\ion{C}{1}     &1657.22 &\phn$ -2.51\pm 1.09 $&\phn$  42\pm  18$ & \phn-- & --\\
					  							
\ion{Al}{2}  &1670.79 &\phn$ -9.48\pm 1.52 $&$ 164\pm  26$&$115\pm30$&220\\
\enddata
\tablenotetext{a}{As presented in figure 5 of Savage et al. 2000}  
\end{deluxetable}

\subsubsection{Emission line strengths}
Once we have taken into account the pronounced absorption for both the \ion{Si}{4}/\ion{O}{4}] blend and \ion{C}{4} with our Gaussian fitting,
these emission features turn out to be the strongest, both with EW\til4\AA.  The various O emission lines are also generally prominent, the more typical lines of
\ion{O}{5} and \ion{O}{4} having EW\til3 and 2\AA\ respectively, whilst even the unusual \ion{O}{3}] feature yields \til1.5\AA.  The weakest line
we measure is that of \ion{C}{3} with only \til0.5\AA. We note that both \ion{N}{5}\lam1240 and \ion{He}{2}\lam1640, common in a large variety
of high excitation objects, are very weak or
entirely absent in our spectrum of \targ.

We also compare our results on the emission line strengths of \targ\ with those published for Her X-1
\citep{ande94,boro96}, Sco X-1 \citep{kall98}, and XTE J1859+226 and XTE J1118+480 \citep{hasw02} based on other \hst\ data.  These systems represent respectively, an
intermediate mass X-ray pulsar, {\em the} ``prototypical'' LMXB, and two rather different soft-X-ray transients (black-hole candidates) in
outburst; hence they
should provide a reasonably diverse context for our study.

Common emission line characteristics of these comparison objects include: (i) very prominent \ion{N}{5} \lam1238, 1242 doublet, (ii) equally strong or stronger
 \ion{C}{4} \lam1548, \lam1550 doublet (except for XTE J1118+480, for which abundance anomalies are postulated), (iii) prominent
 \ion{He}{2}\lam1641 , (iv) similar strength  \ion{Si}{4}/\ion{O}{4}] blend, and (v) usually slightly weaker \ion{O}{5} and \ion{C}{3}\lam1175 (again except for XTE
 J1118+480). Hence, the absence of noticeable \ion{N}{5} and \ion{He}{2} is one distinctly unusual feature of \targ. 
 The lack of \ion{He}{2} provides strong support for the conclusion of  the X-ray abundance analysis of SCH01, namely that the
 white dwarf has no He to
 transfer, but is in fact the core of a more massive initial star, comprised largely of either C-O-Ne or O-Ne-Mg.  The apparent
 reduction in N abundance is then not unexpected. Furthermore, it also appears
 that there is an excess of O in \targ; \ion{O}{4}\lam1342 is very strong here whereas it is only otherwise noted in the case of Sco X-1 (although it may
 also be present in the published spectrum of XTE J1859+226), whilst \ion{O}{3}]  multiplet $\lambda1661-1669$ is not identified in any of these comparison systems.  

In terms of distinguishing between the two possible white dwarf compositions, Mg is the key. As SCH01 discuss, either C-O-Ne or O-Ne-Mg white dwarf
models can account for the high observed Ne and C abundances. In the FUV the only Mg interstellar
absorption line that 
is sometimes seen is \ion{Mg}{2} at 1240\AA\, but in the NUV
there is the \ion{Mg}{2} \lam2796, \lam2803 doublet, which again appears as an interstellar absorption feature but is also found in emission in a
range of astrophysical situations, including quiescent LMXBs \citep{McCl00}.  It is possible that there is weak \lam1240 absorption, whilst
analysis of the \lam2796, \lam2803 doublet (in our NUV spectrum), is rather complex and inconclusive.  The lower spectral resolution of the NUV spectrum makes any
disentangling of emission versus absorption components difficult.  There is marginal evidence for emission, which would in turn imply probable
excess absorption, but equally if we assume no emission the absorption is comparable to what we would expect from interstellar. It would seem
that further higher signal-to-noise/resolution spectra concentrating on this \ion{Mg}{2} doublet would be helpful to derive observational constraints
from this avenue.

\subsection{Emission region dynamics and location}
In terms of the emission line profiles we find two broad categories from our fitting; those consistent with a single emission region and those preferring a 
Doppler pair of emission regions. \ion{C}{3} and  \ion{C}{4} can be represented by emission
from a single region with a broad dispersion in velocities giving a FWHM\til1500\kms\ for the lines, with a negligible shift in line
centers. \ion{Si}{4} is similar (considering the final fit with double-peaked \ion{O}{4}]), as is \ion{O}{3}],  although the FWHM\til900\kms\ and
600\kms\ are somewhat smaller and
100\kms\ velocity shifts were indicated, though for \ion{O}{3}] and possibly for \ion{Si}{4} these may well not be significant. By
contrast, the more highly ionized \ion{O}{5} singlet most likely requires a Doppler pair of emission regions (with $V\sim-1000$ and +600\kms), and similar results are suggested for
the \ion{O}{4} triplet.  Even then the red-shifted components have very large FWHM\til2000\kms, roughly twice that of the blue-shifted components,
which may indicate that the there are in fact multiple red-shifted emission regions. 

Detailed modeling of the various emission regions and their dynamics is beyond the scope of this paper.  However, some physical insights can be
gained by considering results from earlier general studies of UV line formation. It should be noted that all these studies assumed typical
cosmic abundances for the emitting gas, in particular that the heavier elements have trace abundances relative to H and He, which is almost certainly not the case for the
transferred material in \targ. 

\citet{kall82} presented a range of theoretical models for the
ionization structure of gas irradiated by a powerful X-ray source, covering a range of astrophysically relevant parameters.  This enables us to
look at the differing dependence of the various ionic species on the ionization parameter ($\xi=L_X/nR^2$, where $L_X$ is the X-ray irradiating luminosity,
$n$ is the gas density and $R$ is the distance from the emission region to the X-ray source). Their 8 models
include different values of the luminosity, its spectral shape and the gas density. The case of a high density gas ($n=10^{11}{\rm
cm}^{-3}$), illuminated by a 10 keV thermal bremsstrahlung X-ray spectrum with $L_X=10^{37}$\ergsec with line trapping effects included,
is particularly interesting.  Such conditions may well be similar to those in an accretion disc corona/atmosphere above the disc of \targ. The calculations show that the peak of the predicted abundances for our
two groups of emission lines are separated in terms of the value of $\xi$ required. \ion{C}{3}, \ion{Si}{4} and \ion{O}{3} all peak at around
$\log\xi=1.35$,  whereas \ion{O}{4} and \ion{}{5} require a significantly higher $\log\xi\sim1.65$. \ion{C}{4} lies between the two in terms of
$\xi$, but still its abundance is negligible by the time you reach the peak  $\log\xi$ value for the more highly ionized O species. Although
the details will not be exactly applicable to our case, the important point is that such a division in terms of the values of $\xi$ is
plausible.  In consequence, there could be differences in the radial dependence of the emission.  For instance, modeling of the broad emission lines of Her X-1
  (admittedly for a posited disk wind situation) by \cite{chia01} found that the outer radii of the emission region for a given ion were
  systematically smaller for the higher ionization lines.  Similar ``ionization stratification'' was also found from reverberation mapping of
  the broad line regions of AGNs \citep[see e.g.][]{krol91,kori95}.  The effect is probably quite subtle, and in fact specific modeling of the emission lines from X-ray heated accretion discs by \cite{ko94} did not find any clear difference
  in the radial dependence of the line formation between \ion{C}{4} and \ion{O}{5}, although for \ion{Si}{4} the contributions did drop more
  rapidly with decreasing radial location.  They also predicted that overall the greatest contributions to the UV lines should be from the
  outermost radii.

Observationally, the trend in $\xi$ does appear to follow that of the velocity spread/FWHM of our emission lines.  For double-peaked emission
lines the velocity shift of the peaks should be a measure of the Keplerian velocity of the outermost contributing radial annulus
\citep[see][]{smak81,horn86}, whilst for apparently single-peaked lines the HWHM should also be an approximate indicator.  We note that both the single Gaussian and the twin-Gaussian
profiles are merely different approximations to the actual velocity profile one would expect from gas distributed in a Keplerian flow. In fact, the
separation into the two groups is probably a consequence of which is the better approximation for a given velocity spread.

To aid in the comparison of observations with the above models we may also consider the even broader X-ray lines.  SCH01 found that {\em all} these lines had similar
overall width, and hence could be fitted with double-peaked profiles.  In their case, the FWHM of the blue and red-shifted components could be
taken as equal at \til2000--3000\kms, with large velocity shifts for peaks ranging from --1600\kms\ to --2600\kms\ in the blue and 800--1900\kms\ in
the red.  As they comment, these line profiles could also be consistent
with a disc origin.  If one considers the 
velocities of the various X-ray and FUV lines in terms of a Keplerian accretion disc or atmosphere just above (assuming $i=30^\circ$, the upper
limit for a 0.02\msun donor), one would place the outermost contributing radii for \ion{C}{3}, \ion{Si}{4} and \ion{O}{3} at the outer (tidally
truncated) edge ($r_t\simeq2\times10^{10}$cm, where $V\sin i=530$\kms), \ion{C}{4} a little farther in,  \ion{O}{4} and
\ion{O}{5} slightly farther in still and for the X-ray lines from $r\simeq3\times10^{10}$cm to the inner edge (at magnetospheric  corotation, $r_{co}=6.5\times10^{8}$ cm, where $V\sin i=2900$\kms).   But this is where the simple physical picture outlined above appears to break down. As
discussed by SCH01, given the luminosity\footnote{A minimum distance of
  3 kpc was derived by \citet{chak98b} based upon the mass transfer rate implied by the spin-up rate of the pulsar in \targ.} of the source ($L_X=2\times10^{36}(d/3{\rm kpc})^2 \ergsec$), and a gas density of 
$n\sim10^{11}$cm$^{-3}$ requires $R\simgt10^{10}(d/3{\rm kpc})$, comparable to the radius of the outer disc to obtain  $\xi<10^3$ erg cm
s$^{-1}$ and allow formation of  even
the more highly ionized X-ray line emitting species.  For
our FUV lines, with $\xi\sim50$ erg cm s$^{-1}$, the situation becomes even worse, requiring  $R\simgt10^{11}(d/3{\rm kpc})$, almost an order of
magnitude larger than the entire 41.4 min period binary!  One solution might be to assume a higher density; indeed densities exceeding
$10^{16}$cm$^{-3}$ may be possible in the accretion disc proper.  In any case, we suggest that the velocities are more reliably indicating the
radial 
limits of the 
emission locations, and that it is the details of the ionization structure that
need further scrutiny.  Admittedly, there are outstanding issues related to the velocity structure too, notably the asymmetry of the widths of
the Doppler pairs for \ion{O}{5} (or an extra redshifted component), and for all the pairs of {\em both} X-ray and FUV lines the inequality of the red and
blue velocity shifts ($V_{\rm blue}\simeq 2\times V_{\rm red}$).

\subsection{Conclusions}
Analysis of a 32ks time-averaged FUV spectrum of
the ultra-compact X-ray binary, \targ\ has provided further
observational constraints on the nature of the donor star and the
kinematics of the line emission regions in the system. We find evidence
for both a lack of He and N, together with an excess of O in the line 
emitting gas.  Moreover, additional C absorption local to the system is 
probable.
These observations are fully consistent with the X-ray spectroscopic
results of \citet{schu01}, which indicated that the donor is a very
low-mass C-O-Ne or O-Ne-Mg white dwarf.  In addition, we also find 
complicated velocity structures in emission, similar to the Schulz et
al. Doppler pairs of X-ray lines. All the FUV
lines are very broad, with the highest ionization parameter species
exhibiting possibly double-peaked structures.  The interpretation of
these
line profiles as a consequence of emitting regions in a Keplerian disc
about the neutron star is certainly plausible, though the simplest
models may not be adequate. In any case, the combination of high
quality FUV and X-ray spectral line data may provide modelers with
a rich data set for follow-on detailed studies of the disk dynamics
and ionization structure.

\acknowledgments
  Support for this work was provided by NASA
through STScI grant GO-06624.01 and LTSA grant NAG5-7932.


\end{document}